\documentclass[prl,twocolumn,showpacs,floatfix,amsfonts]{revtex4}
\usepackage{graphicx,graphics,color,epsfig} % Include figure files
\usepackage{bm}
\usepackage{amsmath}
\usepackage{amssymb}
\newcommand{\diff}{\mbox{d}}

\def\Rhocomp{^{\vphantom{\vert}}}
\def\Astcomp{^{\vphantom{\ast}}}
\begin{document}

\title{Spin Configurations and Activation Gaps of the Quantum Hall
States in Graphene}
\author{Tapash Chakraborty$^\ddag$ and Pekka Pietil\"ainen$^\dag$}
\affiliation{Department of Physics and Astronomy, University of 
Manitoba, Winnipeg, MB R3T 2N2, Canada \\
$^\dag$Department of Physical Sciences/Theoretical Physics,
  P.O. Box 3000, FIN-90014, University of Oulu, Finland} 

\begin{abstract}
We report on our accurate evaluation of spin polarizations of the
ground state and particle-hole gaps for partially-filled lowest Landau
level, observed in recent experiments on graphene subjected to ultra-high
magnetic fields. We find that inter-electron interactions are important at 
these filling factors, characterized by the non-degenerate ground states
and large particle-hole gaps at infinitely large wave vectors. The gaps 
are the largest for the quantum Hall states in the second Landau level. 
The weak appearence of quantum Hall plateaus in the experiments for certain 
filling factors in the second Landau level however indicates that at these
filling factors, the system has a soft mode at a finite wave vector due 
presumably to the presence of disorder.
\end{abstract}

\maketitle

In recent experimental work, an atomically thin, two-dimensional 
(2D) sheet of graphite, called graphene has been extracted by micromechanical 
cleavage \cite{graphene_2D}. A two-dimensional electron system in 
graphene exhibits many remarkable properties. In the band structure 
calculations where electrons are treated as hopping on a hexagonal lattice 
\cite{wallace}, one finds a unique {\it linear} (relativistic type) 
energy dispersion near the corners of the first Brillouin zone where 
the conduction and valence bands meet. As a consequence, the low-energy 
excitations follow the Dirac-Weyl equations for massless relativistic
particles \cite{ando}. In an external magnetic field, the spectrum
develops into Landau levels, each of which approximately fourfold
degenerate \cite{mcclure,landau_levels}. Recent discovery of the
quantum Hall effect in graphene \cite{qhe_expt,high_field,qhe_theory}
has resulted in intense activities \cite{nat_materials,materials_today,PIC} 
to unravel the electronic properties of graphene that are distinctly different
from the conventional (or, as commonly called the `non-relativistic')
2D electron systems in semiconductor structures.

Graphene has a honeycomb lattice structure of $sp^2$ carbon atoms,
with two atoms A and B per unit cell (i.e., a two-dimensional
triangular Bravais lattice with a basis of two atoms). Each atom is
tied with its three nearest neighbors via strong $\sigma$ bonds
that are in the same plane with angles of 120$^\circ$. The $\pi$
orbit ($2p_z$) of each atom is perpendicular to the plane and
overlaps with the $\pi$ orbitals of the neighboring atoms that
results in the delocalized $\pi$ and $\pi^\ast$ bands. There is
only one electron in each $\pi$ orbit and the Fermi energy is
located between the $\pi$ and $\pi^\ast$ bands \cite{saito}. The
dynamics of electrons in graphene is described by a nearest-neighbor
tight-binding model \cite{wallace} that describes the hopping of
electrons between the $2p_z$ carbon orbitals. The first Brillouin
zone is hexagonal and at two of its inequivalent corners (the K
and K$^\prime$ points) the conduction and valence bands meet.
Graphene is often described as a two valley (K and K$^\prime$)
zero-gap semiconductor. Near these two points (the
so-called Dirac points), the electrons have a relativistic-like
dispersion relation, $\varepsilon\Astcomp_k=\pm\hbar v \vert {\bf k}\vert$
and obey the Dirac-Weyl equations for massless fermions. At the
vanishing gate voltage, the system is half-filled and the Fermi
level lies at the Dirac points.

In graphene, discovery of the Hall conductivity $\sigma_{xy}
=\nu\Rhocomp_G e^2/h = 2(2n+1) e^2/h$, where $n$ is an integer has 
indeed confirmed the Dirac nature of the electrons in that system 
\cite{qhe_expt}. This result has been explained as an outcome of 
four times the quantum Hall conductivity $(n+\frac12)e^2/h$ for 
each Dirac fermion, that is consistent with the four-fold degeneracy 
(two spins and two valleys, or flavors) of the Landau levels for 
the Dirac fermions \cite{qhe_theory}. In this paper, we however
focus our attention to the extreme quantum limit. In the very high 
magnetic field regime (up to 45 Tesla), Zhang et al. \cite{high_field} 
discovered a new set of quantum Hall states at filling factors 
$\nu\Rhocomp_G=0,\pm1,\pm4$. Observation of these states clearly indicated 
that the four-fold degeneracy of the observed states at 
$\nu\Rhocomp_G=\pm4(n+\frac12)$ has been lifted in the presence of 
these ultra-high fields. According to Nomura and MacDonald \cite{nomura},
appearence of the half- and quarter-filled lowest Landau levels are a
consequence of quantum Hall ferromagnetism induced by the electron-electron
interactions. The absence of certain Hall plateaus, such as at 
$\nu\Rhocomp_G=\pm3, \pm5$
are then attributed to the presence of disorder in the system. 

Alicea and Fisher \cite{jason} performed a comprehensive exploration of 
alternative mechanisms for these intermediate quantum Hall states in 
clean and dirty graphene samples. In particular, they looked for 
mechanisms which explicitly could break the inherent symmetries of 
graphene. They started from the lattice scale with a Hubbard type model 
introducing inter- and intra-sublattice interactions to break the 
underlying SU(4) symmetry of the ideal graphene lattice. In the 
ferromagnetic (clean samples) regime they found the system to favor 
flavor polarizations at filling factors $\nu\Rhocomp_G=\pm1$. The 
strength of interactions relative to the Zeeman coupling determines 
whether the system is polarized or unpolarized at $\nu\Rhocomp_G=0$. 
In the higher Landau levels, these authors predicted that skyrmions 
would provide the minimum energy charge excitations. At the filling 
factor $\nu\Rhocomp_G=\pm4$, interactions lead to a spin-polarized 
ground state whereas the states at $\nu\Rhocomp_G=\pm3,\pm5$
break spontaneously the flavor symmetry. In the paramagnetic
(dirty samples) regime the quantum Hall states may be possible
at $\nu\Rhocomp_G=0,\pm1,\pm4$ but not at $\nu\Rhocomp_G=\pm3,\pm5$.

The question of the absence of QH steps at $\nu\Rhocomp_G=\pm3, \pm5$ 
in the experiment has been raised by Fuchs and Lederer \cite{lederer}, 
who, as opposed to the models described above, suggest that 
electron-electron interactions play no role in the formation of 
integer quantum Hall states. Instead they propose that the inversion 
symmetry of graphene is spontaneosly broken by a magnetic field 
dependent Pierls distortion. In the lowest Landau level this leads
to the splitting of the valley degeneracy. This combined
with the Zeeman splitting explains the existense of the
observed Hall plateaus at intermediate filling factors
$\nu\Rhocomp_G=0,\pm1$. In the higher Landau levels,
however, the lattice distortion cannot lift the valley degeneracy.
This leaves only the Zeeman splitting leading to intermediate
plateaus only at $\nu\Rhocomp_G=\pm4,\pm8,\ldots$.

We however believe that inter-electron interactions play an important 
role for the QH states at partially-filled $\nu\Rhocomp_G=1$ that 
do or do not appear as QH steps in the experiments. In this paper, we 
set out to accurately determine the spin configurations and activation 
gaps of various integer filling factors that we expect to shed light 
on the nature of these states as well as provide a better understanding 
of the source of lifting of degeneracy for these states. In conventional 
quantum Hall states at partially filled Landau levels the electron-electron 
Coulomb interaction opens up an excitation gap between the different, 
otherwise degenerate eigenstates of the total spin. In graphene, an 
additional mechanism is necessary to lift also the degeneracy between
the eigenstates of the total flavor pseudospin. Motivated by the lattice 
scale studies by Alicea and Fisher \cite{jason} we introduce into our 
Hamiltonian a term which on the lowest Landau level maps to a charge 
imbalance between the sublattices. We found that the ground states of 
$\nu\Rhocomp_G=\pm1, \pm2, \pm3,$ etc. are non-degenerate, and they have 
large particle-hole gaps at asymptotically large wave vectors -- the 
quasiparticle-quasihole gap for the fractional filling factors 
\cite{fqhe_book,kallin}, which is in fact, the activation gap. Interestingly, 
that gap is much larger for the QH states at the $n=1$ Landau level, 
as compared to that in other Landau levels, which is in line with earlier 
analysis of the effects of Coulomb interactions \cite{goerbig,vadim} 
in graphene. However, in the experiments \cite{high_field} reported as yet,
the absence (or weak presence) of the QH states at $\nu\Rhocomp_G=\pm3, 
\pm5$, etc. would imply that the system has a soft mode at a finite wave 
vector, which might arise due to disorder \cite{jason} in the system.

%\begin{figure}[!ht]
%\centerline{\epsfxsize=8.0cm\epsfbox{Fig_1.eps}}
%\caption[*] {The real and imaginary parts of the energy spectra 
%}
%\end{figure}

In our model the electrons reside on the surface of a torus formed from 
a rectangular cell of sides $L_x$ and $L_y$ by imposing periodic boundary 
conditions in both $x$ and $y$ directions. Choosing the Landau gauge 
$\vec B=\nabla\times(0,Bx,0)^T$, we get the two-component wavefunctions 
$$ \Phi_{nm}^K(x,y)=\frac1{\sqrt2}\left(\begin{array}{c}
-i\,\mbox{sgn}(\lambda)\,\phi_{n-1,m}(x,y) \\
\phi_{nm}(x,y)
\end{array}\right) $$
for the valley $K$ and
$$ \Phi_{nm}^{K'}(x,y)=\frac1{\sqrt2}\left(\begin{array}{c}
\phi_{nm}(x,y) \\ -i\,\mbox{sgn}(\lambda)\phi_{n-1,m}(x,y)
\end{array}\right) $$
for the valley $K'$. Here $n$ is the Landau level index and
$\lambda$ is its energy $\pm\sqrt{2n}$ in units of $\gamma/\ell$,
where $\ell$ is the magnetic length defined as
$ \ell^2=c\hbar/eB$. In the case of the lowest Landau level, 
$n=0$, the components with $n-1$ in the above functions are taken 
to be zero. The component wavefunctions $\phi_{nm}$ are the
Laundau wavefunctions for the non-relativistic systems,
\begin{eqnarray*}
\lefteqn{
\phi_{nm}(x,y)=\left(\frac1{\sqrt\pi L_y\ell 2^nn!}\right)^{1/2}
} \\
&\times&\sum_{q=-\infty}^\infty
e^{i(qL_x-X_m)y/\ell^2}e^{-(x+qL_x-X_m)^2/2\ell^2}\\
&\times& H_n\left((x+qL_x-X_m)/\ell\right)
\end{eqnarray*}
where the centers $X_m$ of the harmonic oscillations are
given by
$$ X_m=\frac{2\pi\ell^2}{L_y}\,m. $$
Due to the periodic boundary conditions, the quantum number
$m$ yielding non-equivalent wavefunctions is restricted to the values
$ m=0,1,\ldots,M-1, $
where $M$ is the degeneracy of the Landau level and also equal
to the number of
flux quanta through the rectangle, i.e.,
$$ M=\frac{L_xL_y}{2\pi\ell^2}. $$

We diagonalize the interaction Hamiltonian
$$H_{ee}=\frac12\sum_{i\not=j=1}^{N_e}\frac{e^2}{\epsilon|\vec r_i-\vec r_j|}$$
in the basis
$$ {\cal B}=
\Bigg\{\left(\begin{array}{c}\Phi^K_{nm} \\ {\bf 0} \end{array}\right),
\left(\begin{array}{c}{\bf 0} \\ \Phi^{K'}_{nm}\end{array}\right)\Bigg|
m=0,1,\ldots,M-1\Bigg\} $$
of the four-component vectors. When appended with the spin degree of
freedom we have effectively a set of eight-component vectors making
the base prohibitively large for practical computations. We
can, however decrease the number of basis states by noting first
that there is no spin dependent term in our Hamiltonian (Zeeman
energies can always be added afterwards) so the total $S_z$ is a
good quantum number. If we fix $S_z$ to its minimum value (0 or 1/2)
the base is significantly reduced but it still allows us to recover all 
possible eigenstates of the total $S^2$. Secondly, because the Coulomb
interaction preserves the quantity $ m=\sum_{i}m_i\,\mbox{mod}\,M, $
performing diagonalization separately for each $m$ reduces the
size of the basis further approximately to the $1/M$-th part.

We are still left with the flavor symmetry associated with the valleys 
$K$ and $K'$ leading to the degenerate ground states. Because on the lowest 
Landau level in the valley $K$ ($K'$)  only the sublattice B (A)
component of the wavefunction differs from zero the electron occupation 
of the valleys maps directly to the occupation of the sublattices. We also 
know that the electrostatic energy is minimized if the electrons reside 
only in one of the sublattices (see \cite{jason}). We utilize this fact by 
introducing the symmetry breaking term
$$ V_{KK}=v\Rhocomp_{KK}\sum_{i\ne j=1}^{N_e}\vec K_i\cdot\vec K_j, $$
where $\vec K$ is the pseudospin operator which is identical to
the spin operators $\vec S$ but works in the flavor (or valley) 
space. Introduction of this term in the Hamiltonian would imply that
either one distributes the charge evenly ($v\Rhocomp_{KK}>0$, $K_i$ and 
$K_j$ antiparallel) between both A and B sublattices or favor one 
of the sublattices ($v\Rhocomp_{KK}<0$, $K_i$ and $K_j$ parallel)
over the other. The latter option applies then only to the lowest 
Landau level. In the higher Landau levels this interpretation is
not valid since the wavefunctions there have equal weight on
both sublattices and the pseudospin operator cannot redistribute the
charges. Therefore, the purpose of the $\vec K\cdot\vec K$ term in 
the present case is only to lift the valley degeneracy. We may also 
speculate (consistently with the experiments) that this kind of 
symmetry-breaking term exists only in the lowest Landau level while 
in the higher levels a new (yet unknown) mechanism 
may operate with increasing magnetic fields.

To construct the interaction Hamiltonian we need to evaluate the
two-body matrix elements
\begin{eqnarray*}
\lefteqn{
A_{n_1m_1,n_2m_2,n_3m_3,n_4m_4}
} \\ &&
=\int\diff\vec r_1\int\diff\vec r_2 \;
\phi_{n_1m_1}^\ast(\vec r_1)\phi_{n_2m_2}^\ast(\vec r_2)
%V(\vec r_1-\vec r_2)\\ &&
\,\frac{e^2}{\epsilon|\vec r_1-\vec r_2|} \\ &&
\times\phi_{n_3m_3}(\vec r_2)\phi_{n_4m_4}(\vec r_1).
\end{eqnarray*}
A straightforward but lengthy algebra leads to the explicit
expression
\begin{eqnarray*}
\lefteqn{
A_{n_1m_1,n_2m_2,n_3m_3,n_4m_4}
} \\
&&={\cal N}_{n_1n_2n_3n_4}{\sum}'_{\alpha,\beta}
e^{i2\pi\alpha(m_1-m_3)/m\Astcomp_{\mbox{\scriptsize max}}}\\
&&\times e^{-\pi(\alpha^2+\beta^2\lambda^2)/
(\lambda m\Astcomp_{\mbox{\scriptsize max}})}
\\
&&\times \frac1{\sqrt{\alpha^2+\lambda^2\beta^2}}
Z_{n_1n_4}(\alpha,\beta)
Z_{n_3n_2}(-\alpha,\beta),
\end{eqnarray*}
where ${\cal N}$ is the normalization constant and we have 
defined the quantities $Z$ as
\begin{eqnarray*}
Z_{nn'}(\alpha,\beta)
&=&\left(\sqrt{\frac{\pi}{2\lambda M}}
(i\alpha+\mbox{sgn}(n-n')\beta\lambda)\right)^{|n-n'|}
\\
&&\times L_{\min(n,n')}^{|n-n'|}\left(\frac{\pi}{\lambda M}
\left(\alpha^2+\beta^2\lambda^2\right)\right).
\end{eqnarray*}
In the summation (with prime) the term with $\alpha=\beta=0$ is excluded and
the index $\beta$ is restricted to the values
$\beta=kM+m_4-m_1;\ k=0,\pm1,\pm2,\ldots$.
For the nonzero matrix elements the quantum numbers $m_i$ must
satisfy the conservation law $m_1+m_2=m_3+m_4\,\mbox{mod}\,M$.

We have evaluated the total spin $S$, the ground state energy
$\varepsilon\Rhocomp_{\rm coul}$ (the Landau level energy is not 
included) and the activation gap (defined as the positive 
discontinuity of the chemical potential $\mu$) \cite{chak_old,fqhe_book},
a measurable quantity
\begin{eqnarray*}
\Delta_g&=&\mu_+-\mu_- \\
&\approx&-2N_e\varepsilon(\nu\Rhocomp_G)+N_e\varepsilon(\nu^+_G)+N_e\varepsilon(\nu^-_G)
\end{eqnarray*}
for Landau levels $n=0,1,2,3$ and for filling factors $\nu\Rhocomp_L=N_e/M$ 
with $N_e$ electrons and $M$ flux quanta in the Landau level at issue. Here
$+ (-)$ refers to the case when a 
particle
%flux quantum
is added to (subtracted from) the
system, and $\varepsilon(\nu\Rhocomp_G)$ is the total energy of the system at the
filling factor $\nu\Rhocomp_G$. The filling factor $\nu\Rhocomp_G$ for graphene 
is related to the conventional (i.e., non-relativistic) filling factor 
\cite{fqhe_book} $\nu\Rhocomp_L+4n$ as 
$$ \nu\Rhocomp_G=\nu\Rhocomp_L+4n-2.$$

\begin{table}
\caption{The Coulomb energy, the activation gap, and the 
ground state spin for four and eight electrons and 
four flux quanta. Here $n$ is the Landau level index,
$\varepsilon_{\rm coul}$, and $\Delta$ are in units of
the Coulomb energy $(e^2/\epsilon l,\, \epsilon=1)$.}

\begin{tabular}[t]{c|c|c|c|c} \hline
\multicolumn{4}{c}{$\nu\Rhocomp_L=4/4$} \\ \hline
$n$&$\nu\Rhocomp_G$&$\varepsilon_{\rm coul}$&$\Delta$& S \\
\hline \\
0&-1&--0.6457&0.5480&2 \\
1&3&--1.0101&1.3256&2 \\
2&7&--0.6587&0.5800&2 \\
3&11&--0.5719&0.3976&2 \\
\hline
\end{tabular}
\hspace{2em}
\begin{tabular}[t]{c|c|c|c|c} \hline
\multicolumn{4}{c}{$\nu\Rhocomp_L=8/4$} \\ \hline
$n$&$\nu\Rhocomp_G$&$\varepsilon_{\rm coul}$&$\Delta$&S \\
\hline \\
0&0&--0.6457&0.5480&0 \\
1&4&--1.0100&1.2632&0 \\
2&8&--0.6559&0.5520&0 \\
3&12&--0.5707&0.4720&0 \\
\hline
\end{tabular}
\end{table}

Our results for various integer filling factors are summarized in
Table I and Table II for systems containing $N_e=4-12$ electrons. 
Clearly, the total spin is $S=N_e/2$ for the 
filling factor $\nu\Rhocomp_L=1$ and $S=0$ for the filling factor 
$\nu\Rhocomp_L=2$ that are independent of the Landau levels. The 
ground state (correlation) energies and the activation gaps are 
presented in Table I for a system with four flux quanta and in 
Table II in the case of six flux quanta. Due to the particle-hole 
symmetry the states with $\nu\Rhocomp_G=4n\pm1$ are equivalent and the ones 
with $\nu\Rhocomp_G=4n+2$ correspond to completely filled Landau levels. 
Therefore they are not shown here.
In our calculations, we have used the value $-0.001$ in units of Coulomb energy
for $v\Rhocomp_{KK}$. It should be pointed out, however, that the
results presented in the tables are not sensitive to the magnitude
of this parameter: in fact, an infinitesimal value would be
 sufficient to yield, e.g., the total spins.

\begin{table}
\caption{Same as in Table~I, but for six and twelve electrons and 
six flux quanta.}

\begin{tabular}[t]{c|c|c|c|c} \hline
\multicolumn{4}{c}{$\nu\Rhocomp_L=6/6$} \\ \hline
$n$&$\nu\Rhocomp_G$&$\varepsilon_{\rm coul}$&$\Delta$&S \\
\hline \\
0&-1&--0.6368&0.6546&3 \\
1&3&--1.1568&1.7256&3 \\
2&7&--0.7947&0.9810&3 \\
3&11&--0.5662&0.5118&3 \\
\hline
\end{tabular}
\hspace{2em}
\begin{tabular}[t]{c|c|c|c|c} \hline
\multicolumn{4}{c}{$\nu\Rhocomp_L=12/6$} \\ \hline
$n$&$\nu\Rhocomp_G$&$\varepsilon_{\rm coul}$&$\Delta$&S \\
\hline \\
0&0&--0.6368&0.6432&0 \\
1&4&--1.1569&1.6908&0 \\
2&8&--0.7947&0.9600&0 \\
3&12&--0.5656&0.5004&0 \\
\hline
\end{tabular}
\end{table}

The non-degenerate ground states (at zero wave vector) for these 
filling factors, as well as large activation gap that appears
at the very large wave vector \cite{kallin,fqhe_book} implies
that the system is incompressible at these filling factors.
Interestingly, as the results indicate, the activation gap is 
always the largest for $n=1$ Landau level. As mentioned above,
this is in line with earlier works \cite{goerbig,vadim}, where electron
correlations were found to be dominant for the $n=1$ Landau
level than for all the other Landau levels, a significant 
shift from the conventional 2D electron systems. In the
present case that would mean that the quantum Hall states
should be most stable for $\nu\Rhocomp_G=\pm3,\pm5, ...$
but in fact, they are conspicuously absent in the reported
experimental results \cite{high_field}. Given the 
incompressible nature of these states for the large
wave vector and for the zero wave vector as found above,
we can only conclude from this apparent contradiction that 
these states must exhibit soft
modes at a finite wave vector, perhaps due to the presence
of disorder \cite{nomura,jason}, or maybe due to a mechanism of the 
type proposed in \cite{lederer} that inhibits the opening of 
the gap at small wave vectors. 

We would like to thank Jason Alicea for helpful discussions. The 
work has been supported by the Canada Research Chair Program, a 
Canadian Foundation for Innovation Grant and the NSERC Discovery 
grant.

\end{document}